\documentclass[prb,twocolumn]{revtex4}

\usepackage{graphicx}
\usepackage{dcolumn}
\usepackage{bm}
\usepackage{amssymb}
\usepackage{dsfont}

\begin{document}

\preprint{APS/123-QED}

\title{On the possibility of level broadening in a quantum dot
due to electrostatic interaction with a gate electrode}

\author{K. M. Indlekofer}
\email{m.indlekofer@fz-juelich.de}
\affiliation{
IBN-1, Center of Nanoelectronic Systems
for Information Technology (CNI), Research Centre J\"ulich GmbH,
D-52425 J\"ulich, Germany }

\date{\today}

\begin{abstract}
In this article, we consider a quantum dot system which is
interacting with a spatially separated metallic gate electrode via
direct Coulomb interaction. Here, the gate electrode is described by
an idealized two-dimensional electron gas. Due to Coulomb scattering
effects, the latter may introduce level broadening to the quantum
dot system.
\end{abstract}

\maketitle

\section{\label{sec:intro}Introduction}

Metallic gate electrodes constitute a common experimental means to
control the energy spectrum of quantum dot structures \cite{Kouwen}.
Electron charges inside the quantum dot typically induce image
charges within the gate electrodes. Obviously, such an electrostatic
influence provides an energetical shift of the quantum dot states.
However, due to Coulomb scattering processes between quantum dot
electrons and electrons inside the dissipative gate electrode, the
dot system in general will be subject to level broadening as well.
In other words, the ``dissipation'' or ``friction'' of gate
electrons becomes visible to the quantum dot, analogous to a Coulomb
drag effect \cite{Rojo}. In this article, we discuss an idealized
model for the modified spectral properties of a quantum dot under
the influence of the Coulomb interaction with a two-dimensional
electron gas (2DEG). Finally, the possible consequences for qubit
systems \cite{qubits,Loss} are discussed.

\section{\label{sec:spec}The model}

\begin{figure}
\includegraphics[width=5cm]{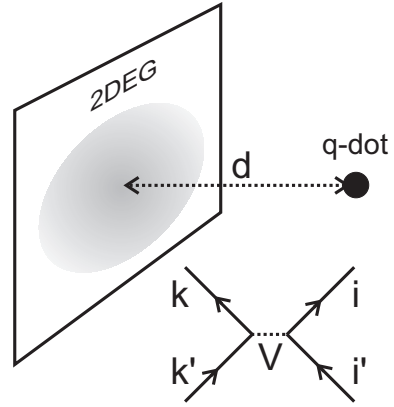}
\caption{\label{fig:fig1} Quantum dot system and gate (2DEG) with
electrostatic interaction ($V$). (The shaded area corresponds to the
induced image charge.) }
\end{figure}

Fig.~\ref{fig:fig1} shows a schematic sketch of the considered
system, consisting of a quantum dot and a 2DEG which represents a
normal metallic gate electrode. In a different context, a similar
system has been discussed by Kato et al. \cite{Kato1,Kato2,Kato3}.
The two spatially separated (i.e. non-overlapping) subsystems
interact with each other via direct electrostatic Coulomb
interaction.

The model Hamiltonian for the interacting quantum dot system and the
Fermi gas of the gate electrode reads as
\begin{equation}
\label{eq:hamop} H=\sum_i E_i a^\dagger_i a_i + \sum_k \epsilon_k
b^\dagger_k b_k +\sum_{i,i',k,k'} V_{ikk'i'}~ a^\dagger_i
b^\dagger_k b_{k'} a_{i'},
\end{equation}
with wavevector indices $k,k'$. (The discrete sum over $k$ can be
considered as the result of periodic boundary conditions for the
Fermi gas states, however, can eventually be replaced by a
$k$-integral in the thermodynamic limit.) $a$ and $b$ are the
quasi-particle annihilation operators for the quantum dot and the
reservoir, respectively. $E$ and $\epsilon$ denote the corresponding
energies, whereas $V$ is the Coulomb matrix for the direct
electrostatic interaction between the quantum dot and the spatially
separated reservoir (i.e., without electron exchange between the two
subsystems due to vanishing overlap of the latter). For simplicity,
intra-dot interaction and the spin degree of freedom will not be
considered in this paper. Furthermore, transitions between the
quantum dot states and the Fermi gas are assumed to be negligible
(due to a sufficient spatial separation).

\section{\label{sec:leh}Fundamental aspects}

In order to understand the main physical effect, we want to focus on
a single quantum dot level $E_0$ in this section. The Hamiltonian thus
reads as
\begin{equation}
\label{eq:hamop0}
H=E_0 a^\dagger a + \sum_k \epsilon_k b^\dagger_k b_k +\sum_{k,k'}
v_{kk'}~ a^\dagger a b^\dagger_k b_{k'},
\end{equation}
where $a\equiv a_0$ and $v_{kk'}\equiv V_{0kk'0}$. The quantum dot
occupation number operator $n=a^{\dagger}a$ has the
eigenvalues 0 and 1. This motivates us to rewrite $H$ as:
\begin{eqnarray}
H&=&
(1-n)\cdot \left(\sum_k \epsilon_k b^\dagger_k b_k\right)\\\nonumber
&&+n\cdot \left(E_0
+\sum_k \tilde{\epsilon_k} \tilde{b}^\dagger_k \tilde{b}_k\right),
\end{eqnarray}
where we have diagonalized the hermitian matrix
\begin{equation}
\hat{\epsilon}_{kk'}=\delta_{kk'}\epsilon_k+v_{kk'}
\end{equation}
by use of a unitary single-particle transformation $U$ such that
\begin{equation}
\hat{\epsilon}_{kk'}=\sum_{j}U_{kj}\tilde{\epsilon}_jU^{\dagger}_{jk'}
\end{equation}
with eigenvalues $\tilde{\epsilon}_k$. Furthermore, we have
introduced the transformed electron operators
\begin{equation}
\label{eq:tildeb}
\tilde{b}_j=\sum_{k}U^{\dagger}_{jk}b_k.
\end{equation}
Since the Hamiltonian $H$ is diagonal with respect to the occupation
number $n$ eigen-subspaces, we obtain
\begin{eqnarray}
H&=&\sum_{k}\epsilon_kn_k \quad\mbox{for}\quad n=0,\\
H&=&\sum_{k}\tilde{\epsilon_k}\tilde{n}_k+E_0 \quad\mbox{for}\quad n=1,
\end{eqnarray}
where $n_k=b^{\dagger}_kb_k$ and
$\tilde{n}_k=\tilde{b}^{\dagger}_k\tilde{b}_k$ are the occupation
number operators for 2DEG states. These two subspace Hamiltonians are
trivial, and the many-body eigenstates can readily be formulated as
Slater determinants:
\begin{eqnarray}
|J\rangle&=&b^{\dagger}_{k(J)_1}\cdots b^{\dagger}_{k(J)_N}|vac\rangle,\\
|K\rangle&=&a^{\dagger}\tilde{b}^{\dagger}_{k(K)_1}\cdots\tilde{b}^{\dagger}_{k(K)_N}|vac\rangle,
\end{eqnarray}
for a Slater determinant $|J\rangle$ with $\langle J|n|J\rangle=0$
and $|K\rangle$ with $\langle K|n|K\rangle=1$. Here, $N$ denotes the
number of 2DEG electrons in the states $|J\rangle$ and $|K\rangle$,
and $(k_1,\ldots,k_N)$ uniquely identifies the occupied
single-particle states within a Slater determinant (with index order
$k_1<\cdots<k_N$). Since we have to consider a grandcanonical
ensemble, all possible Slater determinants with all possible $N$ of
the given type are allowed. As for the eigenenergies of the
eigenstates $|J\rangle$,$|K\rangle$ we obtain
\begin{eqnarray}
E_J&=&E^0_J,\\
E_K&=&E^1_K+E_0,
\end{eqnarray}
with the 2DEG energies
\begin{eqnarray}
\label{eq:e0}
E^0_J&=&\sum_{k}n(J)_k~\epsilon_k,\\
\label{eq:e1}
E^1_K&=&\sum_{k}\tilde{n}(K)_k~\tilde{\epsilon}_k,
\end{eqnarray}
where $n(J)_k$ and $\tilde{n}(K)_k$ denote the occupation numbers in
the Slater determinants for the many-body indices $J$ and $K$,
respectively. One has to note that due to the interaction between
the quantum dot and the gate, the subspaces for $n=0$ and $n=1$ in
general have two different (not trivially overlapping)
single-particle eigenbases for the construction of Slater
determinants as many-body eigenstates. This property will turn out
to be responsible for level broadening in the quantum dot.

We now want to consider the spectral function $A(\omega)$ (density
of states) for the {\it quantum dot} state. This quantity can be
derived from the retarded two-point Green's function of the system
\cite{Mahan,Walecka,Datta}. Within the scope of the many-body
eigenbasis representation, we can directly employ the Lehmann
representation \cite{Walecka} of $A$:
\begin{eqnarray}
A(\omega)&=& -2~\mbox{Im}\left[
\sum_{{J}\atop {\langle J|n|J\rangle=0}}
\sum_{{K}\atop {\langle K|n|K\rangle=1}}
\right.\\\nonumber&&\left.\quad\quad\quad\quad\quad\quad
\frac{\hbar}{\hbar\omega-E_0-(E^1_K-E^0_J)+i\eta}
\right.\\\nonumber&&\left.\quad\quad\quad\quad\quad\quad
\times~(w_J+w_K)~
\right.\\\nonumber&&\quad\quad\quad\quad\quad\quad
\times~\langle J|a|K\rangle\langle K|a^{\dagger}|J\rangle
\Big]\\
&=&
\label{eq:alehdelta}
2\pi\hbar
\sum_{{J}\atop {\langle J|n|J\rangle=0}}
\sum_{{K}\atop {\langle K|n|K\rangle=1}}
\\\nonumber&&\quad\quad\quad\quad\quad\quad
\delta_{\eta}(\hbar\omega-E_0-(E^1_K-E^0_J))
\\\nonumber&&\quad\quad\quad\quad\quad\quad
\times~(w_J+w_K)~
\left|\langle J|a|K\rangle\right|^2,
\end{eqnarray}
with
\begin{equation}
\delta_{\eta}(x)=\frac{1}{\pi}\frac{\eta}{x^2+\eta^2},
\end{equation}
where $\eta\to 0+$ (after the thermodynamic limit for the sum over
2DEG states). From the matrix element of $a$, one can see that only
the combinations $\langle J|n|J\rangle=0$ and $\langle
K|n|K\rangle=1$ with the same $N$ for $J$ and $K$ can provide
non-vanishing terms. $w\geq 0$ are the eigenvalues of the many-body
statistical operator
\begin{equation}
\rho=\sum_{{J}\atop {\langle J|n|J\rangle=0}}\!\!\!
w_J|J\rangle\langle J|
~~+\sum_{{K}\atop {\langle K|n|K\rangle=1}}\!\!\!
w_K|K\rangle\langle K|,
\end{equation}
with grandcanonical equilibrium form
\begin{eqnarray}
w_J&=&\frac{1}{Z}\exp\left(-\beta(E^0_J-\mu\sum_kn(J)_k)\right),\\
w_K&=&\frac{1}{Z}\exp\left(-\beta(E_0-\mu+E^1_K-\mu\sum_k\tilde{n}(K)_k)\right),
\end{eqnarray}
where $\beta=1/(k_BT)$ and $\mu$ denotes the chemical potential, and
$Z$ is the grandcanonical partition function for all $J,K$ such that
$\sum w=1$. This equilibrium condition corresponds to the assumption
of relaxation processes within the gate (which of course are not
explicitly considered in $H$).

The many-body matrix element $\langle J|a|K\rangle$ can be evaluated
by use of the single-particle transformation $U$ in Eq.~(\ref{eq:tildeb}):
\begin{eqnarray}
\langle J|a|K\rangle&=&
\langle vac|
\left(b^{\dagger}_{k(J)_1}\cdots b^{\dagger}_{k(J)_N}\right)^{\dagger}
\\\nonumber &&\quad\quad\quad\quad
\times~\tilde{b}^{\dagger}_{k(K)_1}\cdots\tilde{b}^{\dagger}_{k(K)_N}
|vac\rangle\\
&=&
\langle vac|
b_{k(J)_N}\cdots b_{k(J)_1}
\\\nonumber &&\quad\quad\quad\quad
\times\tilde{b}^{\dagger}_{k(K)_1}\cdots\tilde{b}^{\dagger}_{k(K)_N}
|vac\rangle\\
&=&
\sum_{l_1,\ldots,l_N}U_{l_1k(K)_1}\cdots U_{l_Nk(K)_N}
\\\nonumber &&\quad
\langle vac|
b_{k(J)_N}\cdots b_{k(J)_1}
b^{\dagger}_{l_1}\cdots b^{\dagger}_{l_N}
|vac\rangle.
\end{eqnarray}
By use of $\{b_j,b^{\dagger}_k\}=\delta_{jk}$ and $b_k|vac\rangle=0$
we finally obtain
\begin{eqnarray}
\left|\langle J|a|K\rangle\right|^2&=&
\Big|~ \sum_{l\in Perm(k(J))}
\!\!(-1)^{P(l,k(J))}
\\\nonumber &&\quad\quad\quad\quad
\times~ U_{l_1k(K)_1}\cdots U_{l_Nk(K)_N}
~\Big|^2,
\end{eqnarray}
where $P$ denotes  the parity of the permutation $l$ of the index
set $k(J)$. Hence, $|\langle J|a|K\rangle|^2$ describes the overlap
of Slater determinants for different dot occupation ($J:n=0$ and
$K:n=1$), which becomes non-trivial for $V\neq 0$.

As can be seen from Eq.~(\ref{eq:alehdelta}), the original quantum
dot level at $E_0$ is modulated by 2DEG-induced energy shifts
$E^1_K-E^0_J$. Since $E^1_K-E^0_J$ (see
Eqs.~(\ref{eq:e0}),(\ref{eq:e1})) vanishes for those $J,K$ where
$\left|\langle J|a|K\rangle\right|^2\neq 0$ in the {\it
non-interacting} case, we obtain
$A(\omega)=2\pi\hbar\delta(\hbar\omega-E_0)$ for $V=0$ (note that
the sum over all $w$ is normalized and $U=\mathbf{1}$ w.o.l.g. in
this case). For the {\it interacting} case, however, the sum over
$\delta$-peaks with weights $(w_J+w_K)\left|\langle
J|a|K\rangle\right|^2$ and varying energy shifts $E^1_K-E^0_J$ may
provide an overall shift (i.e., renormalization) of $E_0$ and a
broadening of the $\delta$-peak of the non-interacting case,
depending on $k_BT$ and $\mu$ via $w_J,w_K$. A level broadening
results, if $E^1_K-E^0_J$ has not the same value for all $J,K$ with
$(w_J+w_K)\left|\langle J|a|K\rangle\right|^2\neq 0$. From a
different perspective, the quantum dot electron experiences not only
a classical confinement potential but also the quantum fluctuations
from the non-classical term $\sum_{k,k'}v_{kk'}b^{\dagger}_kb_{k'}$
in $H$ (see Eq.~(\ref{eq:hamop0})). A quantitative estimation of the
expected level broadening will be published elsewhere.

\section{\label{sec:qubit}Possible consequences for charge qubits}
As a consequence, if coupled quantum dot systems are employed as
{\it charge} based qubits that are controlled by external metallic
gate electrodes, the discussed level broadening mechanism might
imply a reduction of the qubit phase coherence time. Even worse, if
multiple, spatially separated gates are used, the electron position
could become ``macroscopically visible'' in terms of measurable
image charges inside the gate electrodes. In simple words, the
spatially resolved multi-gate contacts make the charge position
visible to the rest of the world in terms of detectable charges and
currents through attached cables. This is a typical situation of
marcoscopically distinguishable quantum states, leading to a mixed
state of the reduced density matrix of the qubit system (due to
entanglement with the external experimental setup, coupled via gate
electrodes). Such a mechanism can also lead to a destruction of
entanglement in qubits that are based on the spatial (i.e., charge)
degree of freedom. In order to address these two limitations,
superconducting gate electrodes (in particular, additional screening
gates) might be used as a possible solution. In addition, the latter
gate electrodes might also be bridged internally without any visible
effect to the outside world. The important point is to avoid any
entanglement of the qubits with the environment during quantum
computation. {\it Spin} based qubits \cite{Loss} appear to be more
robust against the discussed gate effects.

\section{\label{sec:sum}Summary}
We have considered a 2D electron reservoir and its influence on a
spatially separated quantum dot system due to a direct Coulomb
interaction. As a main result, the dissipative reservoir may
introduce level broadening to the quantum dot system, which can be
understood in terms of a Coulomb scattering effect (due to
dissipation of gate electrons). Consequences for charge qubit
systems were discussed.

\begin{acknowledgments}
We acknowledge discussions with M. Mo\v{s}ko and A. Bringer.
\end{acknowledgments}

\end{document}